\documentclass[aps,prl,twocolumn,amsmath,nofootinbib,superscriptaddress,amssymb,10pt]{revtex4-1}
\usepackage{graphicx}
\usepackage{ulem}
\usepackage{hyperref}
\hypersetup{bookmarks,
            colorlinks,
           citecolor=red,
            filecolor=blue,
           urlcolor=blue,
           pdfpagemode=None}
\usepackage{txfonts}
\usepackage[mathscr]{euscript}
\usepackage[usenames,dvipsnames]{color}
\usepackage{latexsym}
\renewcommand{\vec}[1]{\boldsymbol{#1}}
\def \n {{\vec n}}

\def \e {\epsilon}
\def \r {{\vec r}}

\def \ve {{\varepsilon}}

\def \D{\Delta}

\def \beq {\begin{eqnarray}}
\def \eeq {\end{eqnarray}}
\def \tn {\textnormal}

\def\e{\eta}
\def\n{\vec{n}}

\def\r{{\bf r}}
\def\R{{\cal R}}

\def\>{\rangle}
\def\<{\langle}

\newcommand{\be}{\begin{equation}}
\newcommand{\ee}{\end{equation}}
\renewcommand{\vec}[1]{\boldsymbol{#1}}
\def \k {{\bf k}}
\def \e {\varepsilon}
\def \r {{\vec r}}

\def \D{\Delta}

\def \beq {\begin{eqnarray}}
\def \eeq {\end{eqnarray}}
\def \tn {\textnormal}
\newcommand{\ba}{\begin{align}}
\newcommand{\ea}{\end{align}}

\newcommand{\bs}{\begin{split}}
\def\sess\end{split}

\begin{document}

\title{Phase transition beneath the superconducting dome in BaFe$_2$(As$_{1-x}$P$_x$)$_2$}
\author{Debanjan Chowdhury}
\affiliation{Department of Physics, Harvard University, Cambridge Massachusetts-02138, U.S.A.}
\author{J. Orenstein}
\affiliation{Department of Physics, University of California, Berkeley, California-94720, U.S.A.}
\author{Subir Sachdev}
\affiliation{Department of Physics, Harvard University, Cambridge Massachusetts-02138, U.S.A.}
\affiliation{Perimeter Institute of Theoretical Physics, Waterloo Ontario-N2L 2Y5, Canada}
\author{T. Senthil}
\affiliation{Department of Physics, Massachusetts Institute of Technology, Cambridge, Massachusetts-02139, U.S.A.}

\begin{abstract}
We present a theory for the large suppression of the superfluid-density, $\rho_s$, in BaFe$_2$(As$_{1-x}$P$_x$)$_2$ in the vicinity of a putative spin-density wave quantum critical point at a P-doping,  $x=x_{c}$. We argue that the transition becomes weakly first-order in the vicinity of $x_{c}$, and disorder induces puddles of superconducting and antiferromagnetic regions at short length-scales; thus the system becomes an electronic micro-emulsion. We propose that frustrated Josephson couplings between the superconducting grains suppress $\rho_s$. In addition, the presence of `normal' quasiparticles at the interface of the frustrated Josephson junctions will give rise to a highly non-trivial feature in the low-frequency response in a narrow vicinity around $x_c$. We propose a number of experiments to test our theory.
\end{abstract}
\maketitle

{\it Introduction.-} An important focus of the study of high temperature superconductivity (SC) has been on the role of antiferromagnetism (AFM) and its relation to SC \cite{Taillefer10}. There is clear evidence across many different families of compounds that SC appears in close proximity to an AFM phase \cite{DS12}; these families include the iron-pnictides, the electron-doped cuprates and the heavy-fermion superconductors. Moreover, the optimal transition temperature ($T_c$) of the SC is often situated where the normal state AFM quantum critical point (QCP) would have been located, in the absence of superconductivity. The experimental detection of the QCP is often challenging in the normal state, and more so in the superconducting state.

Recently, a number of measurements were reported in a member of the pnictide family,  BaFe$_2$(As$_{1-x}$P$_x$)$_2$, as a function of the isovalent P-doping, $x$. The experiments show a phase transition involving onset of spin-density wave (SDW) order in the normal state
above $T_c$, which extrapolates to a $T=0$ SDW QCP (see \cite{Matsudareview} and references therein). These experiments include: ({\it i\/}) a sharp enhancement in the effective mass, $m^*$, upon approaching a critical doping from the overdoped side, as obtained from de Haas-van Alphen oscillations \cite{Matsuda10a} and from the jump in the specific-heat at $T_c$ \cite{walmsley13} , and, ({\it ii\/}) a vanishing Curie-Weiss temperature ($\theta_{CW}$), extracted from the $1/T_1T$  measurements using NMR.

\begin{figure}
\begin{center}
\includegraphics[scale = 0.2]{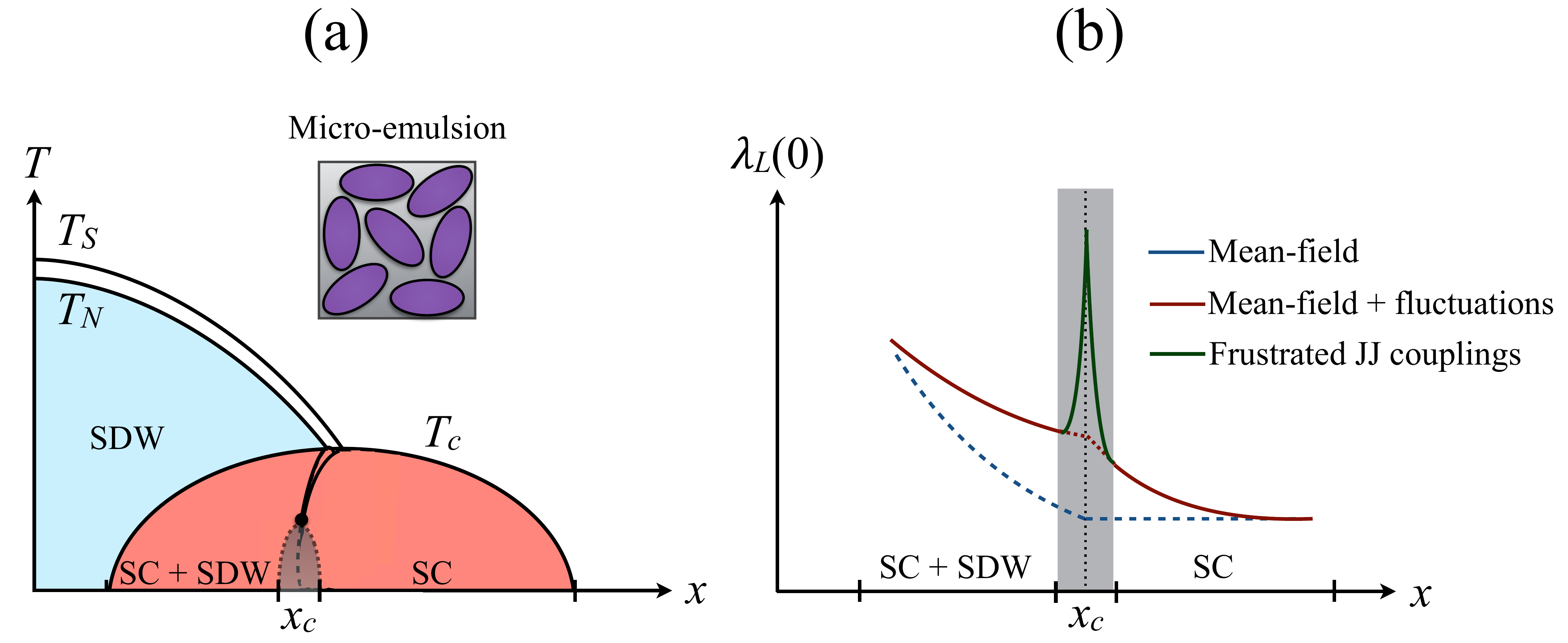}
\end{center}
\caption{(a) A cartoon phase-diagram showing the interplay between SDW and SC phases. The $T_N$ (N$\acute{e}$el temperature) and $T_S$ (structural/nematic transition) lines may survive as continuous transitions inside the SC phase up until the point marked `$\bullet$', below which they become weakly first-order (shown as black dashed line). The grey region depicts the regime where the system develops spatial inhomogeneity due to the presence of disorder. Inset: Emulsion with SC grains (purple) and SDW(+SC) regions (grey). (b) The behavior of $\lambda_L(0)$ as a function of the tuning parameter, $x$, within different scenarios (see text for details).}
\label{ph}
\end{figure}

As we will review below, a number of puzzling results have appeared from experiments investigating whether the SDW QCP actually survives ``under the SC dome.'' Here we propose a resolution of these puzzles by postulating a weakly first-order transition for the onset of SDW order in the presence of SC order (see Fig. \ref{ph}a). Our results are independent of the specific microsopic mechanism responsible for rendering the transition weakly first-order \cite{RFAVC}. It is well known that `random bond' disorder has a strong effect on symmetry-breaking first-order transitions \cite{dis}, and ultimately replaces them with a disorder-induced second order transition in two dimensional systems.  Our main claim is that the inhomogeneities associated with these highly relevant effects of disorder can resolve the experimental puzzles.

The possiblity of a QCP within the SC state was investigated by measurements  \cite{Matsuda12} of
the zero temperature London penetration depth, $\lambda_L(0)\propto1/\sqrt\rho_s$ ($\rho_s\equiv$ superfluid-density), as a function of $x$. A sharp peak in $\lambda_L^2(0)$ was observed at $x=x_c$ and interpreted as evidence for a QCP~\cite{AVC13}. However, this interpretation is at odds with general theoretical considerations \cite{DCSS13} concerning a QCP associated with the onset of SDW order in the presence of a superconductor with gapped quasiparticle excitations \cite{comment1,Nodes}.  These considerations suggest that such systems will display a {\it monotonic\/} variation in $\lambda_L^2(0)$ across the QCP, rather than a sharp peak (see dashed-blue/solid-red curves in Fig.~\ref{ph}b) \cite{DCSS13}.

\begin{figure}
\begin{center}
\includegraphics[scale = 0.22]{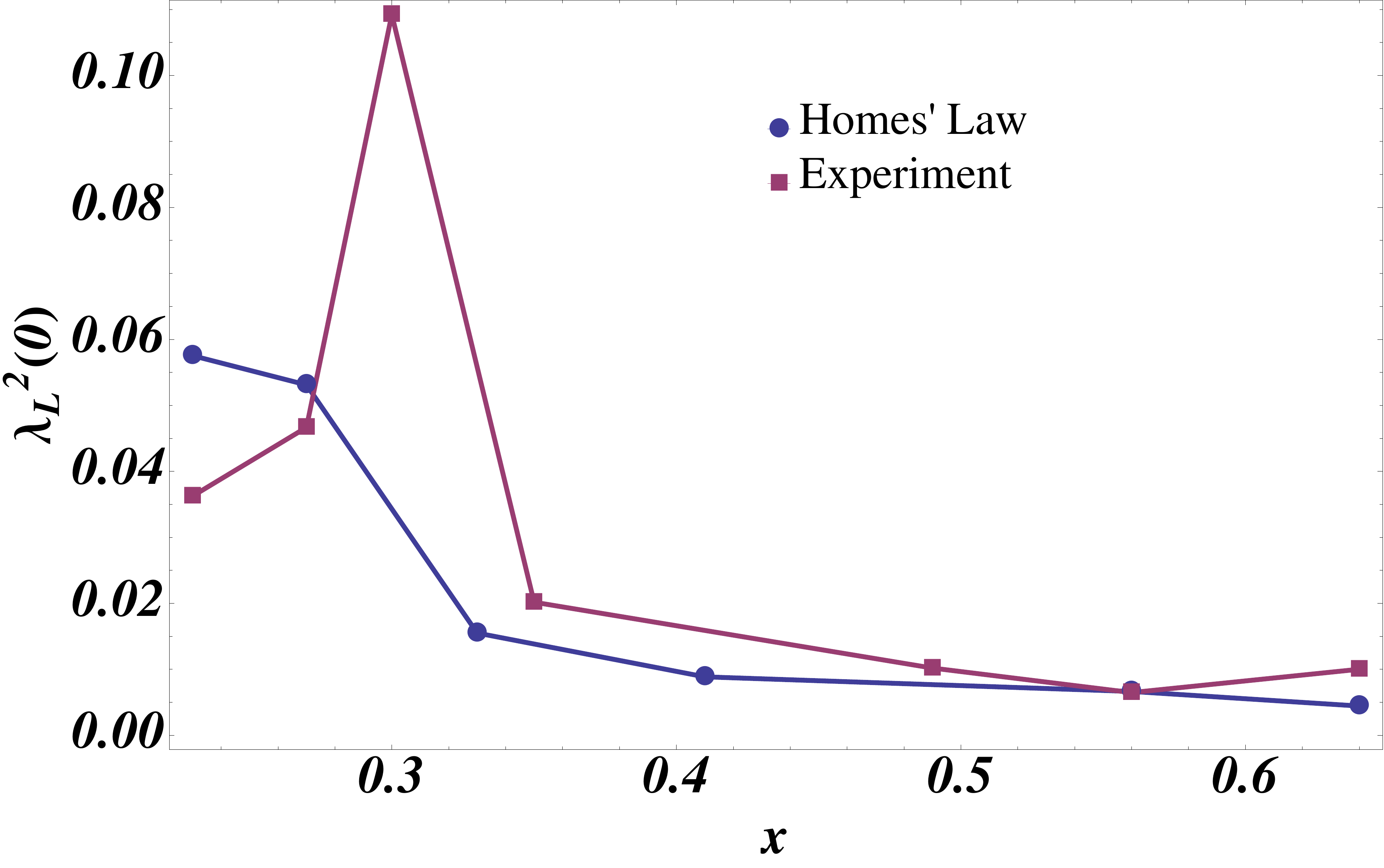}
\end{center}
\caption{A comparison between the experimentally obtained values of $\lambda_L^2(0)$ \cite{Matsuda12} and those deduced from Homes' law. The value of $2\D$ is taken to be $150$cm$^{-1}$ \cite{Matsuda14}, independent of $x$.}
\label{homes}
\end{figure}

As a first step toward resolving this discrepancy, it is useful to place measurements of $\rho_s$ in the context of what is known about the normal state conductivity of the BaFe$_2$(As$_{1-x}$P$_x$)$_2$ system, as these quantities are intimately related through a sum rule. The low temperature superfluid density of a spatially homogeneous superconductor can be estimated from the ``missing area" relation,
\beq
\rho_s\approx \frac{2}{\pi}\Gamma\int_0^{2\Delta/\Gamma} \sigma(z)dz,
\label{homese}
\eeq
where $\Gamma$ is the elastic scattering rate and $z\equiv \omega/\Gamma$. In the dirty limit where $\Delta/\Gamma\ll 1$, the above relation yields Homes' Law \cite{homes}, $\rho_s\approx \sigma(0)\Delta$, whereas in the clean limit $\rho_s=\rho_n$ where $\rho_n$ is the conductivity spectral weight in the normal state. Eqn. \ref{homese} is particularly useful when the normal state resistivity data can reasonably be extrapolated to $T=0$. By combining dc transport data as a function of $x$ \cite{Matsuda10b} and a measurement of 2$\Delta$ from optical conductivity \cite{Matsuda14}, Eq. \ref{homese} provides a lower bound on $\lambda_L^2(0)$ (with the assumption that $\Delta$ is independent of $x$).  Fig.\ref{homes}  shows $\lambda_L^2(0)$ as a function of $x$ obtained under this assumption (details of the procedure are presented as Supplementary Information). The decrease of superfluid density on the underdoped side reflects the growth in residual resistivity that begins as $x$ drops below about 0.33.

The values of $\lambda_L^2(0)$ estimated from Eq. \ref{homese} form a baseline for comparison with the experimental results presented in Ref. \cite{Matsuda12}. On the same graph in Fig. \ref{homes}, we show the experimentally measured $\lambda_L^2(0)$ \cite{Matsuda12}.  The data generally reflect the trend expected from the variation in the residual resistivity, with the exception of the sample with $x=0.3$, in which the condensate spectral weight is suppressed by about 40\% from the Homes' Law estimate. Given the constraints imposed by the sum rule, there are two possible sources of this discrepancy: (\textit{i}) the quasiparticle mass could be renormalized at this value of $x$, corresponding to an intrinsic decrease in $\rho_n$, or, ({\it ii\/}) a considerable fraction of the (unrenormalized) $\rho_n$ could fail to contribute to the low temperature superfluid density. The latter possibility is suggested within the scenario that we develop here.

We analyze the above experiments by assuming a weakly first-order transition \cite{RFAVC}, and argue that the presence of quenched disorder leads to formation of a {\it micro-emulsion} at small scales \cite{dis}. The system consists of SC puddles, where some of the puddles additionally have SDW order (see Fig. \ref{ph}a inset). The SDW(+SC) regions, which have a locally well-developed antiferromagnetic moment but no long-range orientational order, act as barriers between the different SC grains. Upon moving deeper into the ordered side of the transition, the SDW(+SC) regions start to percolate and crossover to a state with long-range SDW order; this is the regime with a microscopically coexistent SC+SDW. As a function of decreasing $x$, the micro-emulsion is therefore a transitional state (shown as grey region in Fig.~\ref{ph}a) between a pure SC and a coexistent SC+SDW. Recent experiments in the vicinity of optimal doping using neutron-scattering and NMR have found results broadly consistent with our proposed phase diagram \cite{dai15}. We note that the granular nature of superconductivity should have no effect on the bulk $T_c$ in the presence of percolating SC channels.

{\it Model.-} When the system is well described in the vicinity of $x_{c}$ by a micro-emulsion as explained above, the phase fluctuations associated with the SC grains (shown as purple regions in Fig.~\ref{ph}a inset), can be modeled by the following effective theory,
\beq
H_{\theta}= - \sum_{a,b}J_{ab}\cos(\theta_a-\theta_b),
\eeq
where $J_{ab}$ represent the Josephson junction (JJ) couplings between grains `$a$' and `$b$'. We have ignored the capacitive contributions.

The Josephson current across the junction will be given by $I_s=J_{ab}\sin(\theta_a-\theta_b)$, and $J_{ab}$ may therefore be interpreted as the lattice version of the local superfluid density, $\rho_s(\r)$, i.e. $\vec{J}_s(\vec{r})=\rho_s(\vec{r})~ \vec{v}_s(\r)$, with $\vec{J}_s(\vec{r}),~\vec{v}_s(\r)$ representing the superfluid-current and velocity respectively. Having a frustrated JJ (also known as a $\pi-$junction) with a negative value of $J_{ab}$ leads to a local suppression in $\rho_s$. Similar ideas have been discussed in the past in a variety of contexts (see Refs. \cite{KivSpi91} for a specific example), though the mechanism considered here will be different. We shall now propose an explicit scenario under which a suppression in $\rho_s$ arises in the vicinity of putative magnetic QCPs, utilizing the SC gap structure in the material under question.

The basic idea is as follows: suppose that the tunneling of electrons between the two grains is mediated by the SDW moment in the intervening region \cite{PWA66},
and is accompanied by a transfer of finite momentum that scatters them from a hole-like to an electron-like pocket. Because the SC gaps on the two pockets have a relative phase-difference of $\pi$, the JJ coupling will be frustrated \cite{Ambegaokar63}.

\begin{figure}
\begin{center}
\includegraphics[scale = 0.22]{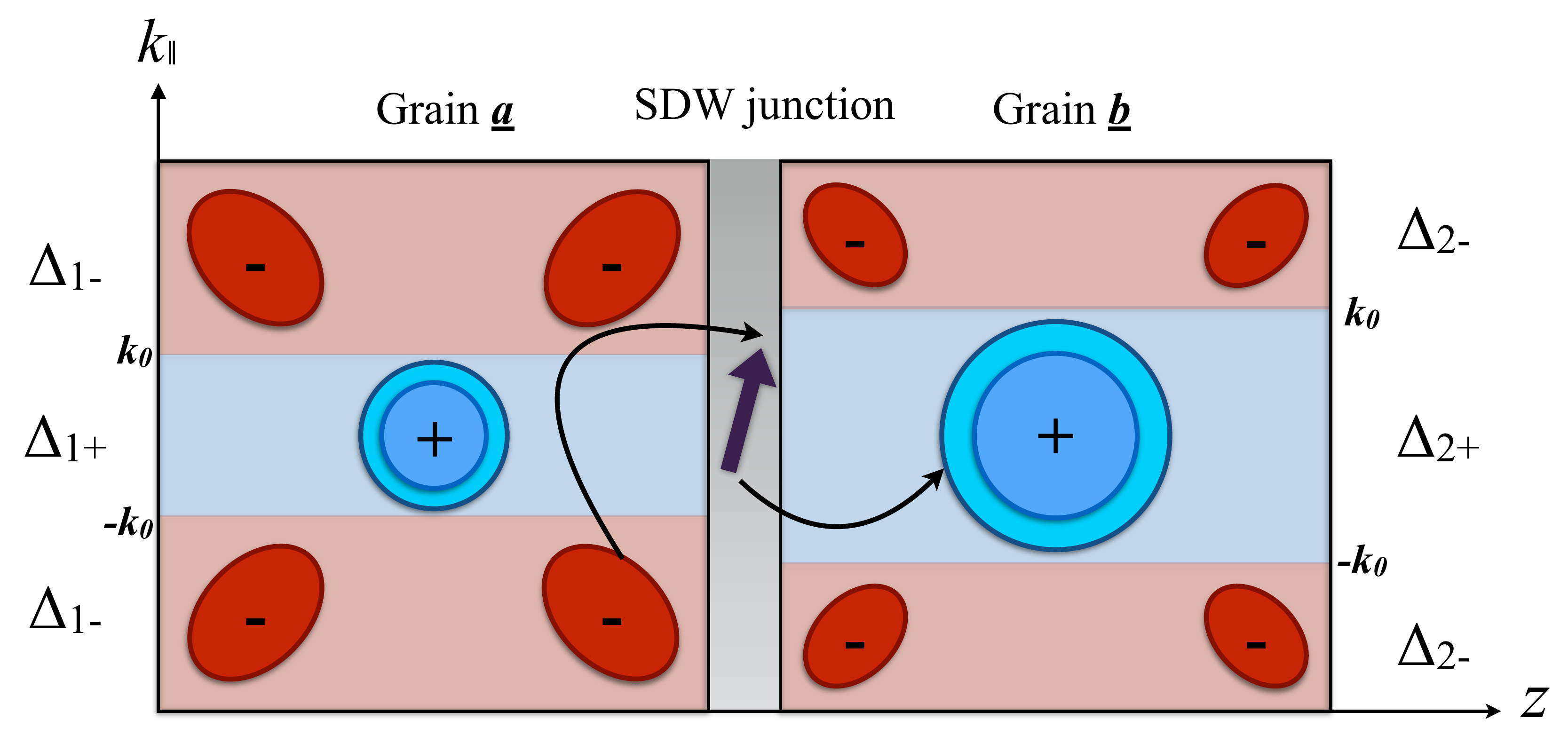}
\end{center}
\caption{A cartoon of a frustrated $\pi-$junction between two superconducting grains with a SDW(+SC) barrier. The SDW moment imparts a finite momentum transfer along the direction of the interface while scattering electrons from the electron (hole) pocket on one grain to the hole (electron) pocket on the other grain.}
\label{JJ}
\end{figure}

Let us first focus on a single grain. In order to capture the multi-band nature of the SCs, we introduce two superconducting order parameters, $\D_i$ with $i=\pm$ to model the $s^\pm$ state on the two pockets. Microscopically, these belong to regions in the grain having different momenta, $\k_\Vert$, parallel to the junction. The gaps are related to the microscopic degrees of freedom \cite{EBNL} via the following relation,
\beq
\D_i(z)=\frac{1}{A}\sum_{\k_\Vert\in \R_i} V_{\k_\Vert,\k'_\Vert}\langle \psi_{\k'_\Vert\uparrow}\psi_{-\k'_\Vert\downarrow}\rangle,
\eeq
where $\psi^\dagger_{\k_{\Vert}\sigma}$ creates an electron at position $z$ with momentum $\k_\Vert$ parallel to the junction and spin $\sigma$. $V_{\k_\Vert,\k'_\Vert}$ is the pairing interaction in the Cooper channel and $z$ is the coordinate perpendicular to the junction with area $A$. The regions $\R_i$ are defined as, $\R_+=\{\k_\Vert| k_0>|\k_\Vert|\}$ and $\R_-=\{\k_\Vert| k_0\leq|\k_\Vert|\}$, where $k_0$ is an arbitrary momentum scale chosen such that $\D_+>0,~\D_-<0 $ (see Fig.~\ref{JJ} for an illustration). We'll assume that such a prescription is valid for each grain, with possibly different values of $k_0$.

Let us then write down a model for the two coupled SC grains with an intervening proximity coupled SDW that has a well developed moment, $\vec\n$. Our notation is as follows: we use $\alpha=a, b$ to denote the grain index and $i=\pm$  to denote the band index within each grain. From now on, we relabel $\k_\Vert$ as $\k$. We introduce the Nambu spinor, $\Psi^{\alpha\dagger}_{i,\k,\sigma}=(\psi_{i,\k,\sigma}^{\dagger\alpha}~\epsilon_{\sigma\sigma'}\psi^\alpha_{i,-\k,\sigma'})$, where now $\psi^{\dagger\alpha}_{i,\k,\sigma}$ creates an electron with momentum $\k$ parallel to the junction and at a position $z$ (label suppressed), which belongs to a region of band ``$i$" within grain ``$\alpha$". The effective Hamiltonian is given by,
\beq
\label{Heff}
H_\tn{eff}&=&H_\D+H_T,\\
H_\D&=&\sum_{\alpha,i,\k} \Psi_{i,\k,\sigma}^{\alpha\dagger}\bigg[\ve_{i\alpha,\k}\hat\tau^z + \D_{i\alpha,\k}\hat\tau^x\bigg]\Psi_{i,\k,\sigma}^{\alpha},\\
H_T&=&g\sum_{k} \vec{n}\cdot\bigg(\Psi^{a\dagger}_{+,\k,\sigma}[\vec\sigma_{\sigma\sigma'}\otimes\hat\tau^0]\Psi_{-,\k,\sigma'}^b \nonumber\\
&&~~~~~~~~~~~+ \Psi^{a\dagger}_{-,\k,\sigma}[\vec\sigma_{\sigma\sigma'}\otimes\hat\tau^0]\Psi_{+,\k,\sigma'}^b \bigg) + \tn{H.c.},
\eeq
where $g$ is the tunneling matrix element, $\hat\tau^i$ $(i=0,x,y,z)$ act in Nambu space and $\hat\sigma^i$  $(i=0,x,y,z)$ act in spin space.

In the above, $H_\D$ corresponds to the bare pairing Hamiltonian written for the $\pm$ bands within each of the two grains. $H_T$ represents the SDW moment mediated hopping of electrons from one grain to the other (represented by the $a, b$ superscripts) and simultaneously scattering from one band to the other (represented by the $\pm$ subscripts). Therefore, $\vec\n$ imparts a finite momentum (along the interface) to the electrons when it scatters them from the electron (hole) pocket on one grain to the hole (electron) pocket on the other grain (shown as the black arrows in Fig.~\ref{JJ}).

{\it Results.-} Using the Ambegaokar-Baratoff relation \cite{Ambegaokar63}, we can write the Josephson coupling (at $T=0$) between the two grains as,
\beq
J_{ab}= \frac{g^2\langle\n^2\rangle}{\pi^2}\bigg[ \sum_{\ell\in a, \ell'\in b} \D_\ell \D_{\ell'}\int_0^\infty\frac{d\e_\ell}{E_{\ell}}\int_0^\infty\frac{d\e_{\ell'}}{E_{\ell'}}\frac{1}{E_\ell+E_{\ell'}}\bigg]
\eeq
where $E_\ell^2=\e_\ell^2+\D_\ell^2$ and $\ell,~\ell'$ represent the band indices on the different grains. Since $\D_\ell\D_{\ell'}<0$, the coupling $J_{ab}<0$. Note that the specific nature of the frustrated tunneling arises from the same spin-fluctuation mediated mechanism that is predominantly responsible for the $s^\pm-$ pairing symmetry \cite{DS12}. However,  there will also be a direct tunneling term (not included in Eqn. \ref{Heff}) in the Hamiltonian, which does not scatter the electrons from one pocket to the other, as they hop across the junction. The contribution to the JJ coupling from this term will be unfrustrated (i.e. $J_{ab}>0$).

The ratio of the tunneling amplitudes in the two different channels is non-universal and depends on various microscopic details. In particular, the emulsion is associated with a distribution of Josephson-couplings, ${\cal{P}}(J)$, with a mean coupling strength, $\langle J\rangle=\bar{J}$. 
 If a substantial fraction of the JJ couplings become negative due to the mechanism proposed above, $\bar{J}$ will be small, and the superfluid density will be suppressed (see green curve in Fig.\ref{ph}b).

\begin{figure}[ht!]
\begin{center}
\includegraphics[scale = 0.27]{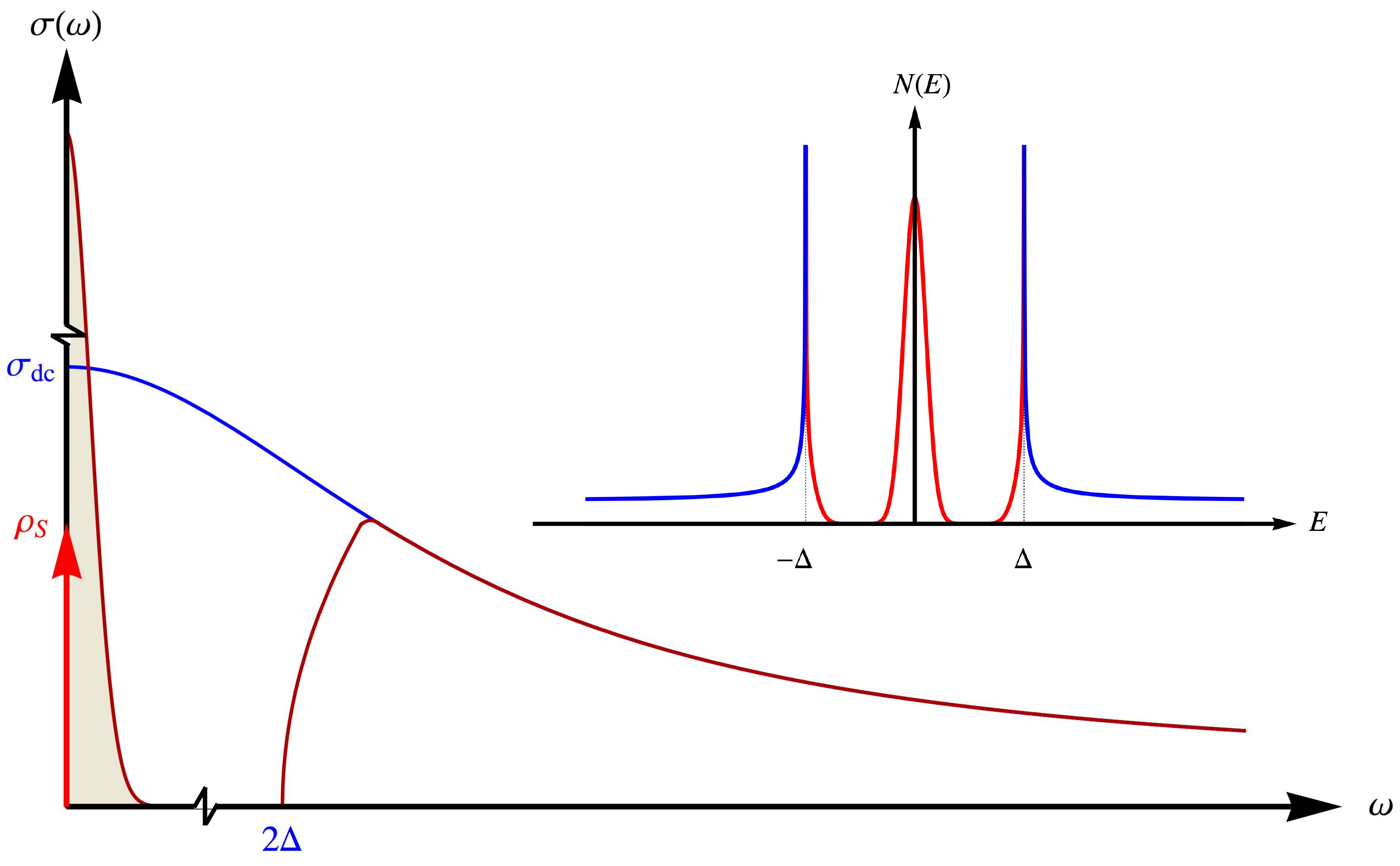}
\end{center}
\caption{Plot of the optical conductivity, $\sigma(\omega)$, for a Drude metal (blue curve) and a superconductor with a large number of sub-gap excitations (red curve). The shaded region corresponds to the displaced spectral weight from the superfluid-density, $\rho_s$ (red arrow). Inset: Density of states, $N(E)$ vs. $E$, for a conventional gapped superconductor (blue curve). A superconductor with a large number of subgap states has a finite $N(E)$ below $\D$ (red curve).}
\label{sigw}
\end{figure}

We now propose a resolution as to the fate of the uncondensed spectral weight (highlighted in Fig. \ref{homes}), which can potentially be tested by measurements of the low frequency optical conductivity. Frustrated $\pi-$junctions host gapless states at the interface between the two grains \cite{RJCR76,hutanaka}, giving rise to a finite density of states around zero energy (see Fig~\ref{sigw} inset). As a result of the gapless `normal'-fluid component at the interface, a fraction $f$ of the spectral weight will be displaced from the superfluid-density to non-zero frequencies (shaded region in Fig.~\ref{sigw}). Given that the weight of the condensate is proportional to $\bar{J}(1-f)$, the 40\% suppression in $\rho_s$ for BaFe$_2$(As$_{1-x}$P$_x$)$_2$ in the vicinity of the putative QCP corresponds to $f\sim 0.6$. 

Our proposed optical conductivity, $\sigma(\omega)$, in the vicinity of penetration depth anomaly is shown in in Fig~\ref{sigw}. The spectrum shows clearly that the connection between normal state conductivity and superfluid density implied by Eq. \ref{homes} will break down. In particular, $\sigma_{\tn{dc}}$ (which is a property of the normal state), could vary monotonically with isovalent-doping across $x_c$, while the abundance of low-energy excitations in the immediate vicinity of $x_c$ would give rise to a non-monotonic variation in the superfluid density. This allows for an unusual way of rearranging spectral weight in the {\it superconducting} state below the gap, without violating optical sum-rules.

The above scenario will give rise to a number of interesting low temperature thermodynamic and transport properties, as we now discuss. First of all, there should be a striking enhancement in the low-temperature thermal conductivity and specific-heat, as a function of $x$ in the narrow vicinity of $x_c$, due to the `normal'-component. It is important to recall that this material has loop-like nodes on the electron-pockets \cite{Nodes}. However, the geometry of the electron-pockets and the magnitude of the gap do not change substantially in the vicinity of $x_c$, and therefore it is unlikely that the contribution to the above quantities from the nodal-quasiparticles will have a drastic modificiation. It should therefore be relatively straightforward to disentangle the contribution arising from the nodal versus the `normal' quasiparticles. Studying the NMR-spectra as a function of decreasing temperature (across $T_c$) and down to sufficiently low temperatures in the vicinity of $x_c$ should also reveal the spatial inhomogeneity associated with the SDW regions. A large residual density of states in the superconducting state has been detected at a particular P-doping via the power-law temperature dependence of $1/T_1\sim T$ \cite{ishida10}. Within our scenario, there should be a striking enhancement in this quantity as a function of doping around $x_c$. Finally, we note that a promising direction for future studies would be to measure the magnetic-field distribution due to the propagating currents in the emulsion using NV-based magnetometers \cite{AY13}.

{\it Discussion.-}  The theoretical study in this paper was motivated by a number of remarkable experiments carried out in BaFe$_2$(As$_{1-x}$P$_x$)$_2$, as a function of $x$ in the normal and superconducting phases. Our primary objective was to provide an explanation for the striking enhancement of the London penetration depth in the vicinity of a putative SDW QCP in the SC state. We developed a scenario based on the idea that true SDW criticality is masked by a weak first-order phase transition in the superconducting state at $T=0$. In this picture, quenched disorder naturally gives rise to an {\it emulsion} at small length scales with puddles of SC and SDW(+SC). It is then, in principle, possible for SDW moments at the interface of the SC grains to generate frustrated Josephson couplings, which deplete the local superfluid-density. Our proposed scenario naturally calls for a number of experimental tests that should be carried out in the near future, which should directly look for both the spatial inhomogeneities associated with the emulsion \cite{curro15}, and probe the gapless excitations using thermodynamic probes, as explained above.

In addition to experiments on BaFe$_2$(As$_{1-x}$P$_x$)$_2$, it should be important to further investigate the contrasting behavior of the electron-doped system, Ba(Fe$_{1-x}$Co$_x$)$_2$As$_2$, where $\lambda_L(0)$ behaves monotonically as a function of $x$ across the putative QCP \cite{Gordon10}. Electron-doping leads to significantly higher amounts of disorder compared to the isovalently-doped case, and would therefore lead to puddles with typically much smaller size \cite{Curro13}. Our proposed mechanism for the strong suppression of the superfluid-density in the isovalently-doped material relies on the existence of an emulsion with puddles of appreciable size, in the presence of an optimal amount of disorder. A comparison of the NMR spectra in the narrow vicinity of the putative QCP in the electron and isovalently doped materials would shed light on these microscopic differences between the two families.

Finally, though we have hypothesized that the SDW onset transition {\it inside} the SC is, in the absence of disorder,  a weak first order transition, we emphasize that the normal state properties are consistent with the presence of a ``hidden" QCP around optimal doping \cite{walmsley13,Matsuda10a, IF14}. It is plausible that in the normal state, different experimental techniques are probing the critical fluctuations associated with not one, but distinct QCPs as a function of $x$. For instance, $m^*$ extracted from high-field quantum oscillations is dominated by the vicinity of `hot-spots', where quasiparticles are strongly damped due to coupling to the SDW fluctuations \cite{TS14}. On the other hand, strong critical fluctuations associated with the nematic order-parameter \cite{FCS14}, that couple to the entire Fermi-surface, would dominate $m^*$ extracted at zero-field from the jump in the specific heat at $T_c$.

{\it Acknowledgements.-} We thank A. Carrington, A. Chubukov, N. Curro, J.C. Davis, R. Fernandes, K. Ishida, M.-H. Julien, S. Kivelson, Y. Matsuda, A. Millis and A. Vishwanath for useful discussions. We thank K. Hashimoto and Y. Matsuda for providing us with the data shown in Fig.\ref{homes}. DC is supported by the Harvard-GSAS Merit Fellowship and acknowledges the ``Boulder summer school for condensed matter physics - Modern aspects of Superconductivity", where some preliminary ideas for this work were formulated. DC and SS were supported by NSF under Grant DMR-1360789, the Templeton foundation, and MURI Grant W911NF-14-1-0003 from ARO. TS was supported by Department of Energy DESC-8739- ER46872, and partially by a Simons Investigator award from the Simons Foundation. JO acknowledges the Office of Basic Energy Sciences, Materials Sciences and Engineering Division, of the U.S. Department of Energy under Contract No. DE-AC02-05CH11231 for support. Part of this work was completed when JO was visiting MIT as a Moore Visitor supported by grant GBMF4303. Research at Perimeter Institute is supported by the Government of Canada through Industry Canada and by the Province of Ontario through the Ministry of Research and Innovation.

\subsection{Supplementary Material}
\section{Estimate of $\lambda_L^2(0)$ from Homes' law}
In this supplementary material, we compare the value of the penetration depth obtained from experiments \cite{Matsuda12} with the prediction from Homes' law; for the latter, we use a combination of the experimental data obtained from optical-conductivity and dc transport. For each value of the doping ($x$), we estimate the (approximate) dc resistivity ($\rho_{xx}$) by extrapolating the curves to $T=0$, from the transport data in fig.1(b) of Ref.\cite{Matsuda10b}.  

We estimate the value of $2\D$, where $\Delta$ is the superconducting gap, from the data for optical conductivity in the superconducting state, as shown in fig. 3(b) of Ref. \cite{Matsuda14}. Since $T_c$ remains relatively unchanged as a function of $x$ in the vicinity of optimal doping, we assume $\D$ to be independent of $x$ such that $2\D\approx150 $cm$^{-1}(=2.827\times10^{13}$s$^{-1})$. Then, in the dirty limit,
\beq
\rho_s = \frac{4}{\pi} \sigma_{\tn{dc}}\Delta.
\eeq

In order to obtain the penetration depth, we need to restore various dimensionful constants such that, 
\beq
\lambda_L^2(0)=\frac{c^2\ve_0}{\rho_s},
\eeq
where $c(=3\times10^{8}$ m/s) is the speed of light and $\ve_0(=8.85\times10^{-12}$ F/m; 1~F=1~$\Omega^{-1}$s) is the permitivity of free space. The values obtained are shown in the table below and have been presented in fig. {\color{red} 2} of the main text, along with a comparison to the experimental data \cite{Matsuda12}.
\begin{center}
\begin{tabular}{ ||c||c|c|c||} 
\hline\hline
$x$ &~ $\rho_{xx}$ (from Ref. \cite{Matsuda10b})& $\lambda_L^2(0)$\\ 
 & ($\mu\Omega$~cm)  &($\mu$m$^2$)  \\
\hline\hline
\hline
0.23 & 130  & 0.057 \\
\hline
0.27 & 120  & 0.053 \\
\hline
0.33 & 35  & 0.015 \\
\hline
0.41 & 20  & 0.009\\
\hline
0.56 & 15  & 0.007\\
\hline
0.64 & 10  & 0.004\\
\hline
\hline
\end{tabular}
\label{data}
\end{center}

\begin{thebibliography}{99}
\bibitem{comment1}There is reason to believe that there are accidental (loop-like) nodes on the electron-pockets in the s$^\pm$ state in this particular material (see refs. \cite{Nodes}); this feature doesn't affect most of the qualitative features of the computation as long as the nodes do not coincide with the SDW ``hot-spots".
\end{thebibliography}
\end{document}